\documentclass[journal, draftclsnofoot, 12pt, onecolumn]{IEEEtran}
\usepackage{lineno,hyperref}
\usepackage{amsmath,amssymb}
\usepackage{float}
\usepackage{graphicx}
\usepackage{booktabs}
\usepackage{makecell}
\usepackage{multirow}
\usepackage{tabu,bm}
\usepackage{color}
\usepackage{subfigure}
\usepackage{lipsum,mwe,cuted}
\usepackage{booktabs}
\usepackage{stfloats}
\usepackage{algorithm}  
\usepackage{algorithmic}
\usepackage{url}
\usepackage{cite}
\usepackage{enumitem} 
\usepackage{epstopdf}
\usepackage[style=base]{caption}
\newtheorem{definition}{Definition}
\newtheorem{theorem}{Theorem}
\hypersetup{hidelinks}

\begin{document}

\title{Semantic Communications: Principles and Challenges\\ \large(Invited Paper)}

\author{Zhijin Qin,~\IEEEmembership{Senior Member,~IEEE,} Xiaoming Tao,~\IEEEmembership{Senior Member,~IEEE,} Jianhua~Lu,~\IEEEmembership{Fellow,~IEEE,} Wen Tong,~\IEEEmembership{Fellow,~IEEE,} \\and Geoffrey Ye Li,~\IEEEmembership{Fellow,~IEEE}
\thanks{Zhijin Qin is with the School of Electronic Engineering and Computer Science, Queen Mary University of London, London E1 4NS, U.K. (e-mail: z.qin@qmul.ac.uk)}
\thanks{Xiaoming Tao and Jianhua Lu are with the Department of Electronic Engineering and also with the Beijing National Research Center for Information Science and Technology, Tsinghua University, Beijing, China. (e-mail: taoxm@tsinghua.edu.cn, lhh-dee@mail.tsinghua.edu.cn) }
\thanks{Wen Tong is with Wireless Technology Labs, Huawei Technologies, Ottawa, Canada (e-mail: tongwen@huawei.com)}
\thanks{Geoffrey~Ye~Li is with the School of Electrical and Electronic Engineering, Imperial College London, London SW7 2AZ, U.K. (e-mail: geoffrey.li@imperial.ac.uk)}}

\maketitle
\begin{abstract}
Semantic communication,  regarded as the breakthrough  beyond the Shannon paradigm, aims at the successful transmission of semantic information conveyed by the source rather than  the accurate reception of each single symbol or bit regardless of its meaning. This article provides an overview on semantic communications. After a brief review of Shannon information theory, we discuss semantic communications with theory, framework, and system design enabled by deep learning. Different from the symbol/bit error rate  used for measuring  conventional communication systems,  performance metrics for semantic communications are also discussed. The article concludes with several open questions in semantic communications.
\end{abstract}

\section{Introduction}
Around 70 years ago,  Weaver~\cite{weaver1953recent} categorized communications into three levels: transmission of symbols, semantic exchange, and effects of semantic exchange. The first level communications, transmission of symbols, has been well studied and  delivered in conventional communication systems, which are approaching to the Shannon capacity limit. However, in many situations,  the ultimate goal of communications is to exchange  semantic information, such as  natural languages, while the communication  medium, such as optical fiber, electromagnetic wave, and cable, can only transmit physical signals. Recently, semantic communication\footnote{In this article, semantic communications refer to  Level 2 and Level 3, also called task/goal-oriented communications.} has attracted extensive attention from industry and academia~\cite{DBLP:journals/corr/abs-2012-08285,tong2021challenges}, and has been identified  as a core challenge for the sixth generation (6G) of wireless networks.

\begin{figure}[t!]
\centering
\includegraphics[width=5.8in]{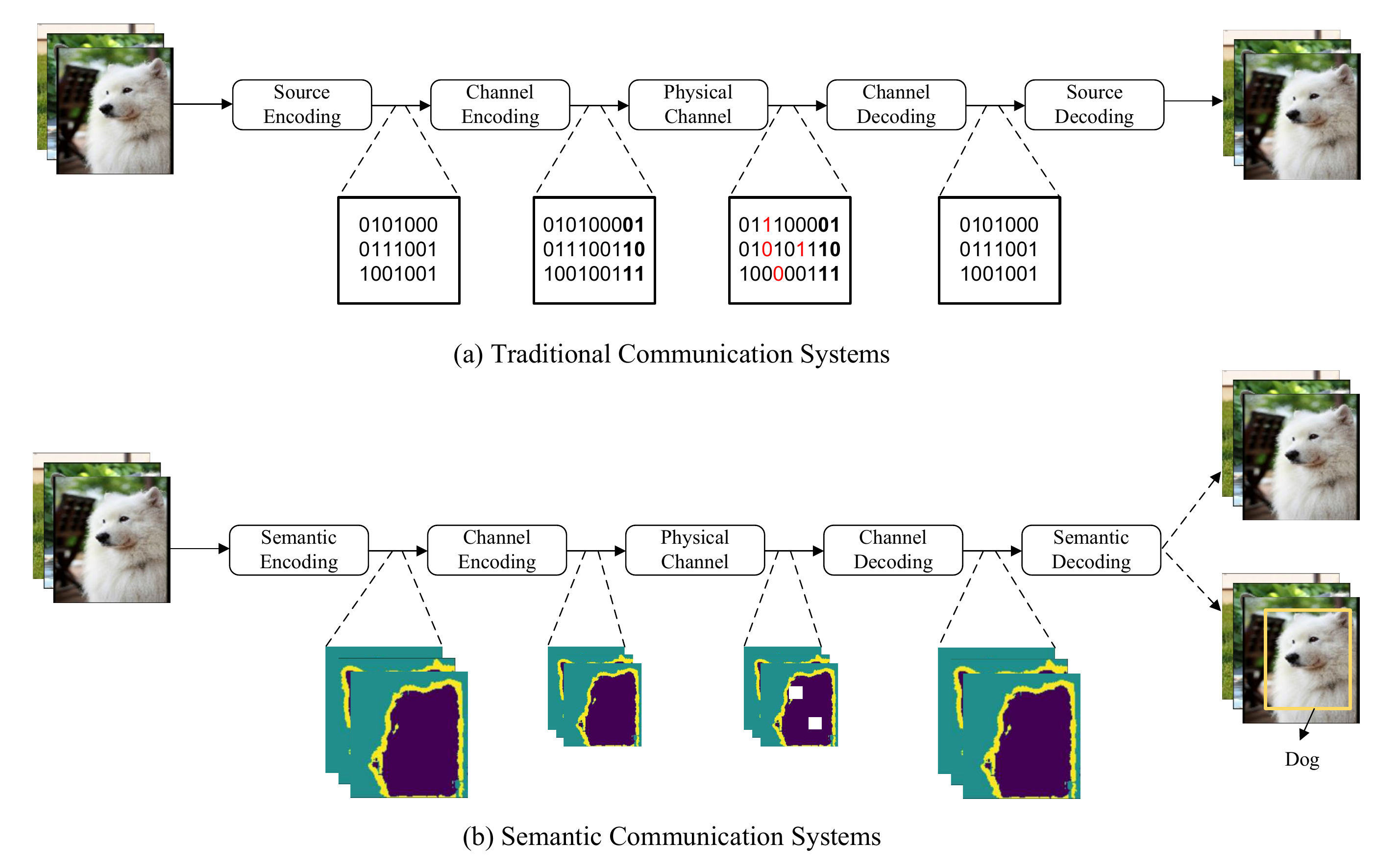}
\caption{The illustration of a semantic communication system for object recognition.}
\label{fig_illu}
\end{figure}

In contrast to the Shannon paradigm,  semantic communications only transmit  necessary information relevant to the specific  task at the receiver~\cite{weaver1953recent}, which leads to a truly intelligent  system with significant reduction in data traffic. Fig.~\ref{fig_illu} demonstrates the concept of semantic communications, where the transmission task is image recognition. Instead of transmitting  bit sequences representing the whole image, a semantic transmitter extracts  features relevant to recognize the object, i.e., dog, in the source. The irrelevant information, such as image background, will be omitted  to minimize the transmitted data without degrading performance. As a result, the demands on energy and wireless resources will be lowered significantly, leading to a more sustainable communication network.

For many applications in the age of 6G and artificial intelligence (AI), the agent, such as smart terminals, robot, and smart surveillance, is able to understand the scene and  executes the instruction automatically. Hence, the core of the task-oriented semantic communication is  deep semantic level fidelity rather than shallow bit-level accuracy. Such a semantic communication framework will be widely used in Industrial Internet, smart transportation, video conference, online education,  augmented reality (AR), and virtual reality (VR), to name a few.

Since the masterpiece from Shannon~\cite{weaver1953recent}, the remarkable progress has been made in understanding the mathematical foundation of  symbol transmission without considering semantics of the transmitted  symbols or bits. Bar-Hillel and Carnap~\cite{carnap1952outline} revisited the bypassed semantic problem in Shannon’s work, and offered a preliminary definition of semantic information. Bao~\textit{et al.}~\cite{bao2011towards} clarified the concepts of semantic noise and semantic channel. A semantic communication framework~\cite{semantic2018game} was proposed to minimize  semantic errors. These pioneering works are  based on  logical probability and are mainly designed for textual processing. Due to the lack of a general mathematical model to represent semantics, the development of semantic communications is still in its infancy after seven decades since it was first introduced.

Recent advancements in deep learning (DL) and its applications, such as natural language processing (NLP), speech recognition, and computer vision   provide  significant insights on developing semantic communications~\cite{tong2021challenges,QINDL:2019}. Hwang~\cite{Hwang:2011}  discussed  intelligence transmission in the  analysis and design of communication systems. Moreover, Chattopadhyay \textit{et al.}~\cite{chattopadhyay2020quantifying} quantified  semantic entropy and the complexity for semantic compression. Joint source-channel coding (JSCC) schemes~\cite{gold2018,bourtsoulatze2019deep} were proposed to capture and transmit  semantic features, in which the semantic receiver executes the corresponding actions directly rather than recovering the source messages. More recently, a series of semantic communication frameworks have been developed \cite{xie2020deep,Wen2021JSAC,xie2020lite,xie2021taskoriented} for multimodal data transmission, which have attracted extensive attention.

So far, there have been several tutorials and surveys on semantic communications. Tong~\emph{et al.}~\cite{tong2021challenges} identified two semantic communication related critical challenges faced by AI and 6G,  including its mathematical foundations and the system  design. Kalfa \emph{et al.}~\cite{KALFA2021103134} discussed  semantic transformations of different sources for popular tasks in the field, and presented the semantic communication system design for  different types of sources. Strinati~\emph{et al.}~\cite{Strinati:2020}  indicated the role of  semantic communications in 6G.  Lan~\emph{et al.}~\cite{Lan2021WhatIS} classified semantic communications into human-to-human (Level 2), human-to-machine (Level~2 and Level 3), and machine-to-machine (Level 3) communications. Various potential applications of semantic communications are  presented. Furthermore, many researchers have dedicated to designing new frameworks~\cite{Shi:2020,ZHANG2021,Pappas:2020,uysal2021semantic} for semantic communications in the format of terse and forceful magazine articles. By 
summarizing the highly related works, they   serve as a good start  to step into the area.

In this article, we will provide a comprehensive overview on  principles and challenges of semantic communications. We first review Shannon information theory and summarize the development of semantic theory. After clarifying the critical difference between conventional and semantic communications, we  introduce  principles, frameworks, and performance metrics of semantic communications. Next, we present the developments of DL-enabled semantic communications for multimodal data transmission, including text, image, and audio. We  conclude the article with research challenges to pave the pathway to  semantic communications. This tutorial article tries to provide insights on answering the following common questions:
\begin{itemize}
    \item \textit{How to understand the semantic meaning of bit sequences?}
    \item \textit{Where is the gain from in semantic communications?} 
    \item \textit{Is there a theoretic limitation for semantic communication systems?} 

\end{itemize}

This article will present readers a clear picture of  semantic communications. The rest of the article is structured as follows. Section II compares conventional and semantic communication systems and theories. Section III presents  semantic communication system components, semantic noise, and performance metrics. The recent advancements on DL-enabled semantic communication systems for transmitting  multimodal data  are discussed in Section IV. This article  concludes with open questions in Section V.

\section{From Information Theory to Semantic Theory}
In this section, we  first discuss the critical difference between conventional and semantic communications. Then we  briefly introduce  information theory and semantic  theory.

\subsection{Difference between Conventional and Semantic Communications}
\begin{figure}[t]
\centering
\includegraphics[width=6.6in]{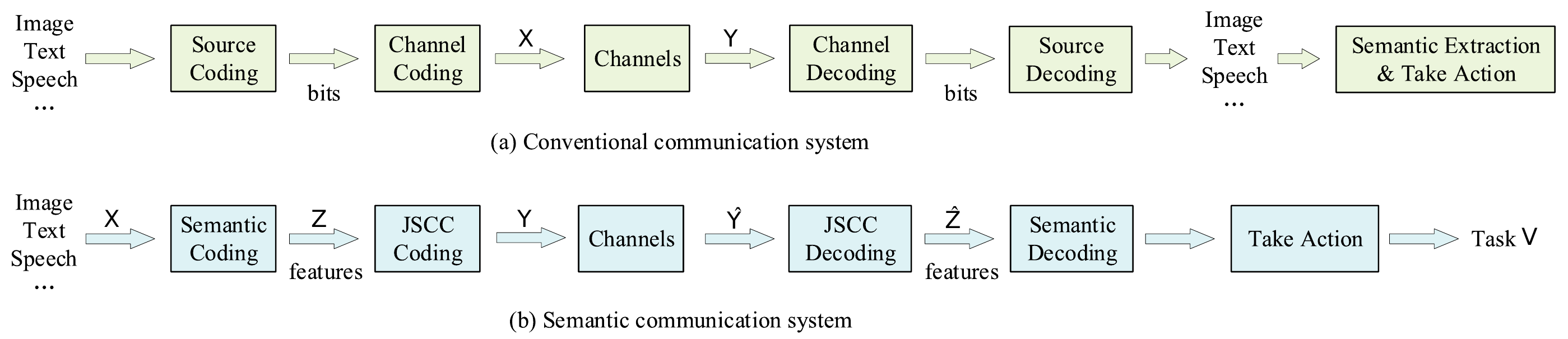}
\caption{Comparison of conventional and semantic communication systems.}
\label{fig_com}
\end{figure}

Since research in semantic communications is still in its preliminary stage,  there is no consistent definition of semantic communications yet.  In Fig.~\ref{fig_com}, we compare  conventional and semantic communications. In a conventional communication system, the source is converted into  bit sequences to process. At the receiver, the bit sequence representing the source is recovered accurately. In a conventional communication system, the bit/symbol transmission rate  is bounded by Shannon capacity. Semantic communications transmit semantic meaning of the source. One of the critical difference is the introduction of semantic coding, which captures the semantic features, depending on tasks or actions to be executed at the receiver. Only those semantic features will be transmitted, which  reduces the required communication resources significantly. Tasks at the receiver could be data reconstruction or some more intelligent tasks, such as image classification and language translation. Note that in semantic communications, data are not processed at the bit level, but the semantic level. Semantic communications could be described by semantic theory.

The following  starts with a brief review of  information theory. We  then present  semantic theory developed in  past decades though it is not well established yet.

\subsection{Information Theory}
In 1948, Shannon  introduced the concept of information entropy~\cite{weaver1953recent}, which exploits uncertainty to measure the information content in terms of bits.

\begin{definition}\label{def-1}
Given the source, $X\in\{{x_1},{x_2}, \cdots ,{x_n}\}$, with probabilities $\left \{ p(x_i)\right\}_{i=1}^{n}$, the source entropy measures the average number of bits  per symbol to be reconstructed without loss, which is defined as
\begin{equation}
    H(X) =  - \sum\limits_{i = 1}^n {{p(x_i)}{{\log }_2}} {p(x_i)}.
\end{equation}
\end{definition}

\begin{theorem}\label{def-2}
For transmission over noisy channels, described by $p(y_j|x_i)$,  channel capacity  is given by
\begin{equation}
   C = \mathop {\max }\limits_{p(x)} I\left( {X;Y} \right),
\end{equation}
where $ I(X;Y)=H(X)-H(X|Y) $ is the mutual information between the input, $X$, and the output, $Y$, of the  channel,  and  $H(X|Y)=- \sum\limits_{j = 1} {p(y_j)}\sum\limits_{i = 1} {{p(x_i|y_i)}{{\log }_2}} {p(x_i|y_i)}$ is the conditional entropy of $X$ for a given $Y$ as shown in Fig.~\ref{fig_com}(a).
\end{theorem}
With the channel capacity, Shannon further developed the source-channel coding theorem.
\begin{theorem}
\textit{If $x_i$ for $i=1,...,n$, satisfies the asymptotic equipartition property (AEP) and $H(X) \le C$, there exists a source-channel code with probability of error $p( {\hat x }_i\ne x_i) \to 0$. Conversely, there will be a positive probability of error if $H(X) > C$.}
\end{theorem}

Conventional communication systems are based on  Shannon's \textit{separation theorem} including two stages: i) compress source data into its most efficient form; and ii) map the sequence of source coding into channel coding.

\textit{Theorem 1} provides an upper bound for distortion-less transmission. For a given distortion $D^{*}$, the minimum transmission information rate $R$ can be described by  rate distortion theory, also known as the lossy source coding theorem.
\begin{theorem}
\textit{For a given maximum average distortion $D^{*}$, the rate distortion function $R(D^{*})$ is the lower bound of the transmission bit-rate~\cite{Blanchart2011FastLM}
\begin{equation}
    R\left( {{D^{\rm{*}}}} \right) = \mathop {\min }\limits_{D \le {D^{\rm{*}}}} I\left( {X;Y} \right),
    \label{Q32}
\end{equation}
where $D = \sum\nolimits_{x,y} {p\left( x \right)} p\left( {y\left| x \right.} \right)d\left( {x,y} \right)$ is the distortion, $d(x, y)$ is the distortion metric with $d(x, y)=0$ if $x=y$. As $  R\left( {{D^{\rm{*}}}} \right)=\mathop {\min }\limits_{D \le {D^{\rm{*}}}} I\left( {X;Y} \right) \leq \mathop {\min }\limits_{D=0 } I\left( {X;Y} \right)=H(X)$, $R\left( {{D^{\rm{*}}}} \right) \le H(X)$.}
\end{theorem}

Information theory is a rich subject nowadays and it is impossible to introduce it within a couple of pages. The above are only theorems and definitions that are highly related to the semantic theory in the subsequent discussions. 

\subsection{Semantic Theory}
Entropy  in Shannon information theory measures the information content by  the uncertainty of the source. However,  how to measure the amount of semantic information, or the importance of information,  for a specific transmission task is yet to be determined. The transmission task refers to the task, such as classification, recognition, and device configuration, to be  performed at the receiver after the source information is received in conventional communications.
\begin{definition}\label{def-3}
Given a transmission task, $V$, the semantic information, $Z$, is the information relevant to $V$ in the source, $X$.
\end{definition}
The uncertainty in $Z$ is less than that in $X$, which indicates the following relationship:
\begin{equation}\label{eq2}
    H(V) \le H(Z) \le H(X).
\end{equation}

The semantic information, $Z$, is extracted from $X$, which can be regarded as the lossy compression of $X$. However, from the view of $V$, $Z$ is the lossless compression of $X$ since the task, $V$, could be fully served by it. With different tasks, the required semantic representation, $Z$, will be different as shown in Fig.~\ref{fig_com}.(b).

With the transmission task, $V$, it is possible to measure the importance of information, say semantic information. For example, for image classification tasks, the receiver is only interested in the objects in an image rather than the original image. Therefore, the objects are considered as essential information while the others are non-essential. Similarly, for text transmission, the receiver requires the meaning of text instead of  lossless text recovery. 

\subsubsection{\bf{Semantic Entropy}}

In the past decades, researchers have  worked hard to find a way to quantify semantic entropy by following the path of information entropy developed by Shannon. However, it is still an active research area with huge space to investigate.

Based on  logical probability, many different definitions of semantic entropy  have been developed.  Carnap and Bar-Hillel~\cite{carnap1952outline} measured the semantic information by  the \textit{degree of confirmation}, which is expressed as
\begin{equation}
    H({\cal H}, e) = -\log c({\cal H}, e),
\end{equation}
where $c({\cal H}, e)$ is defined as the degree of confirmation of hypothesis $\cal H$ on the evidence, $e$. For example,  $\cal H$ could be a new message and $e$ could refer to the knowledge. Bao \textit{et al.} \cite{bao2011towards} defined  semantic entropy of a message or sentence $s$ as 
\begin{equation}
H(s)=-{\log}_2(m(s)),
\label{Q4}
\end{equation}
where the  logical probability of $s$ is given by
\begin{equation}
 m\left( s \right) = \frac{{p \left( {{{\cal{W}}_s}} \right)}}{{p \left( \cal{W} \right)}} = \frac{{\sum\limits_{w \in {\cal{W}},w \models s}^{} {p \left( w \right)} }}{{\sum\limits_{w \in \cal{W}}^{} {p \left( w \right)} }}.
 \label{Q3}
\end{equation}
Here, $\cal{W}$ is the symbol set of a classical source, $\models$ is the proposition satisfaction relation, and ${{\cal{W}}_s} = \left\{ {\left. {w \in \cal{W}} \right|w \models s} \right\}$ is the sets of  models for $s$, i.e., the space in which $s$ is true. Here, $p \left( w \right)$ is the probability  of $w$. If there is no background available, ${\sum\limits_{w \in W}^{} {p \left( w \right)} }=1$. The conditional entropy with background knowledge is also defined by extending the above definition.

In fuzzy systems, semantic entropy is defined by introducing the concept of \textit{matching degree} with the \textit{membership degree}~\cite{Liu_Fuzzy:2019}. Membership degree is a concept in fuzzy set theory and is usually  difficult to measure analytically. Therefore, it is  defined manually by following expert intuition and experience. Liu \textit{et al.}~\cite{Liu_Fuzzy:2019} defined  $\varsigma $ as a semantic concept, which could be treated as the transmission task, and ${\mu _\varsigma }\left( {{X}} \right)$ as the membership  degree for each $X \in {\cal{X}}$,where ${\cal{X}}$ is the set of $X$. For the class, $C_j$,  the matching degree, $ {D_j}\left( \varsigma  \right) = \frac{{\sum\limits_{X \in {{\cal{X}}_{{C_j}}}}^{} {{\mu _\varsigma }\left( X \right)} }}{{\sum\limits_{X \in {\cal{X}}}^{} {{\mu _\varsigma }\left( X \right)} }}$, characterizes the semantic entropy of $X$ on the concept,  $\varsigma $. Note that the definition of matching degree shares  similarity as that defined in~\eqref{Q3}. With the matching degree,  semantic entropy on class $C_j$ is defined as
\begin{equation}
    {H_{{C_j}}}\left( \varsigma  \right) =  - {D_j}\left( \varsigma  \right){\log _2}{D_j}\left( \varsigma  \right).
\end{equation}
The overall semantic entropy over $\cal{X}$ could be obtained by summing up that of all classes. The basic properties of semantic entropy defined above are similar to those of information entropy. The difference is  the membership degree. It is related to the semantic concept or the transmission task, which characterizes the semantic information. 

The above definitions of semantic entropy assume that there exists a way to measure  semantic information. All  theorems are developed based on the assumption of available semantic representation without providing the specific approach to quantifying  semantic information. A group of  statisticians~\cite{chattopadhyay2020quantifying} is working on an active project to develop an information-theoretic framework for quantifying semantic information content in multimodal data, where semantic entropy is defined as the minimum number of semantic enquires about the source $X$, whose answers are sufficient to predict the transmission task, $V$. By doing so, the  approach to finding  semantic entropy becomes  to find the minimal representation of $X$ for serving the task, $V$. However, it  is still under investigation and how to apply such a framework to practical scenarios is to be clarified. 

\subsubsection{\bf{Semantic Channel}} For communication over a noisy channel, the received message is usually with distortion. From the Shannon theorem, such errors caused by distortion can be measured by bit-error rate (BER) or symbol-error rate (SER), which is the engineering problem. From the semantic view, such errors can be measured by  semantic mismatch.  Bao \textit{et al.} \cite{bao2011towards} introduced two kinds of semantic errors from logic probability, 1) Unsoundness: the sent message is true but the received message is false, 2) Incompleteness: the sent message is false but the received message is true. However, there is yet a definition of the semantic error/noise, which will be further discussed in Section III.A.

\subsubsection{\bf{Semantic Channel Capacity}}
In addition to~\eqref{Q4}, Bao \textit{et al.} \cite{bao2011towards} further developed the following theorem for semantic channel capacity, which can be regarded as counterpart of \textit{Theorem~\ref{def-2}}.
\begin{theorem}
\textit{The semantic channel capacity of a discrete memoryless channel is expressed as}
\begin{equation}
    {C_s} = \mathop {\sup }\limits_{p\left( {Z|X} \right)} \left\{ {I\left( {X;V} \right) - H\left( {\left. Z \right|X} \right) + \overline {{H_S}\left( V \right)} } \right\},
\end{equation}
\textit{where ${I\left( {X;V} \right)}$ is the mutual information between the source, $X$, and the transmission task, $V$. ${p\left( {\left. Z \right|X} \right)}$ is the conditional probabilistic distribution that refers to a semantic coding strategy with the source, $X$, encoded into its semantic representation, $Z$, and ${H\left( {\left. Z \right|X} \right)}$ means the semantic ambiguity of the coding. ${\overline {{H_S}\left( V \right)}=-\sum p\left(v\right) H\left(v\right) }$, $v\in V$, is the average logical information of the received messages for the task V.}
\end{theorem}

Note that higher ${H\left({Z|X}\right)}$ means higher semantic ambiguity caused by the semantic coding while higher ${{H_S}\left( V \right)}$ leads to strong  ability for the receiver to interpret received messages. The semantic channel capacity could be either higher or lower than the Shannon channel capacity $I\left( {X;V} \right)$, dependent on the semantic coding strategy  and the receiver's ability to interpret  received messages. 

Two cases are provide here for better understanding. Given the source sentence ``She parked Jame's car on the ground floor of the building, which has 13 floors with 120 sqm on each floor and is called \textit{Smith Building} due to the creator, William Smith.'' The receiver want to know where is Jame's car.
\begin{itemize}
    \item \textbf{Case 1}: $\overline {{H_S}\left( V \right)} - {H\left({Z|X}\right)}> 0$, which means that the receiver can handle the semantic ambiguity. The source sentence can be compressed as ``the ground floor of \textit{Smith Building}.'' Compared with the source sentence, the semantic ambiguity is higher, which means that ${H\left({Z|X}\right)}$ increases. However, the receiver can answer the question with the received sentence, thus the receiver can handle the semantic ambiguity. The semantic compression can achieve higher transmission rate. $C_s$ is higher than Shannon capacity in this case.
    \item \textbf{Case 2}: $\overline {{H_S}\left( V \right)} - {H\left({Z|X}\right)}< 0$, which means that the receiver cannot solve the semantic ambiguity. The source sentence can be compressed as ``She parked Jame's car on the building.'' The receiver cannot find the car based on the received sentence, thus the receiver cannot handle the semantic ambiguity, in which $C_s$ is lower than Shannon capacity.
\end{itemize}

\subsubsection{\bf{Semantic Rate Distortion and Information Bottleneck}}
Similar to~\eqref{Q32},  the rate distortion in semantic communication system is formulated as~\cite{Liu:2021:Rate} 
\begin{equation}\label{DistortionRate}
    R\left( {{D_s},{D_a}} \right) = \min I\left( {Z;\hat X,\hat Z} \right),
\end{equation}
where ${D_s}$ is the semantic distortion between  source, $X$, and  recovered information,  $\hat X$, at the receiver, and ${D_a}$ is the distortion between  semantic representation, $ Z$, and   received  semantic representation, $\hat Z$. Note that \eqref{DistortionRate} considers the distortion caused by both semantic compression and channel noise.

Information bottleneck is an approach to finding the optimal tradeoff between compression and accuracy, which is to solve the following problem~\cite{tishby2000information}
\begin{equation}
    \mathop {\min }\limits_{p\left( Z|X \right)} I\left( {X; Z} \right) - \beta I\left( {V;Z} \right),
\end{equation}
where V is the desired semantic representation. As its extension, Sana \textit{et al.}~\cite{sana2021learning} designed a new loss function as
\begin{equation}
\mathcal{L} = \underbrace {{I }\left( {Z;X} \right)}_\text{Compression} - \underbrace {\left( {1 + \alpha } \right){I}\left( {Z;\hat Z} \right)}_{\text{Mutual}{\kern 1pt} {\kern 1pt} \text{information}} + \underbrace {\beta  {KL\left( X, \hat Z \right)} }_\text{Inference},
\label{Loss3}
\end{equation}
where $\alpha$ and $\beta$ are the parameters to adjust the weights of the mutual information term and the inference term. The compression term represents the average number of bits required for $X$. The inference term is the Kullback-Leibler (KL) divergence between the  posterior probability at the encoder, $X$, and the one captured by the decoder, $\hat Z$. Note that the upper bound of \eqref{Loss3} is
\begin{equation}\label{Xie_loss}
\mathcal{L}={\mathcal{L}_{CE}} - \alpha {I}\left( {Z;\hat Z} \right), 
\end{equation}
which is the loss function  designed in \cite{xie2020deep} to be detailed in Section VI.

Though it is not yet possible to quantify semantic communication systems as Shannon did for conventional communication systems, understanding the above concepts could provide us important insights, especially  on the loss function design for DL-enabled semantic communications.

\section{Components, Semantic Noise, and Performance Metrics}
As we can see from Section II, semantic  theory is still in its infancy. But it does not prevent us from developing practical semantic communication systems.  This section introduces main components, semantic noise, and performance metrics for semantic communication systems. 

\subsection{Semantic Communication System Components}
\begin{figure}[t]
\centering
\includegraphics[width=5.2in]{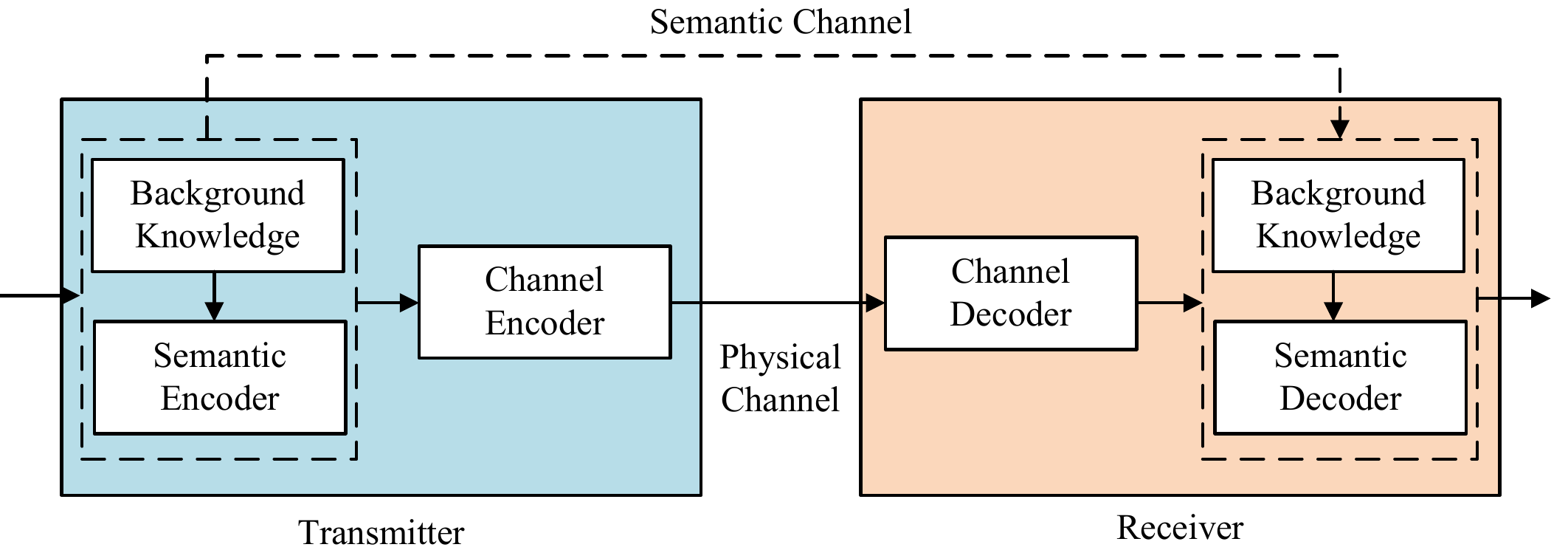}
\caption{The main components in a semantic communication system.}
\label{fig_semsys}
\end{figure}

As shown in Fig.~\ref{fig_semsys}, a semantic communication system includes semantic level and transmission level. The semantic level  addresses semantic information processing  to obtain  semantic representation, which is performed by the semantic encoder and decoder. Here, the semantic information refers to that useful for serving the intelligent tasks at the receiver. The transmission level  guarantees the successful reception of symbols at the receiver after going through the transmission medium, which are normally proceeded by the channel encoder and decoder.  The  semantic transmitter and   receiver  are equipped with certain background knowledge to facilitate  semantic feature extraction, where the background knowledge could be different for various applications.

Similar to the general structure shown  in Fig.~\ref{fig_semsys}, a task-oriented semantic signal processing framework has been proposed~\cite{KALFA2021103134}. Moreover, semantic-aware active sampling~\cite{Pappas:2020,uysal2021semantic} allows each smart device to control its traffic, in which  samples are generated if the sampler is  triggered for serving a specific task.

Note that there are two types of channels to deal with in semantic communication systems. The first type of channels are the physical channels, which introduce channel impairments, such as noise, fading, and inter-symbol interference,  to  the transmitted symbols. In the past, the majority of efforts on wireless communications have been made to combat the physical channel impairments. The second type of channels are the semantic channels, which could be contaminated by semantic  noise caused by misunderstanding, interpretation errors, or disturbance in the estimated information.

\begin{figure}[t]
\centering
\includegraphics[width=6in]{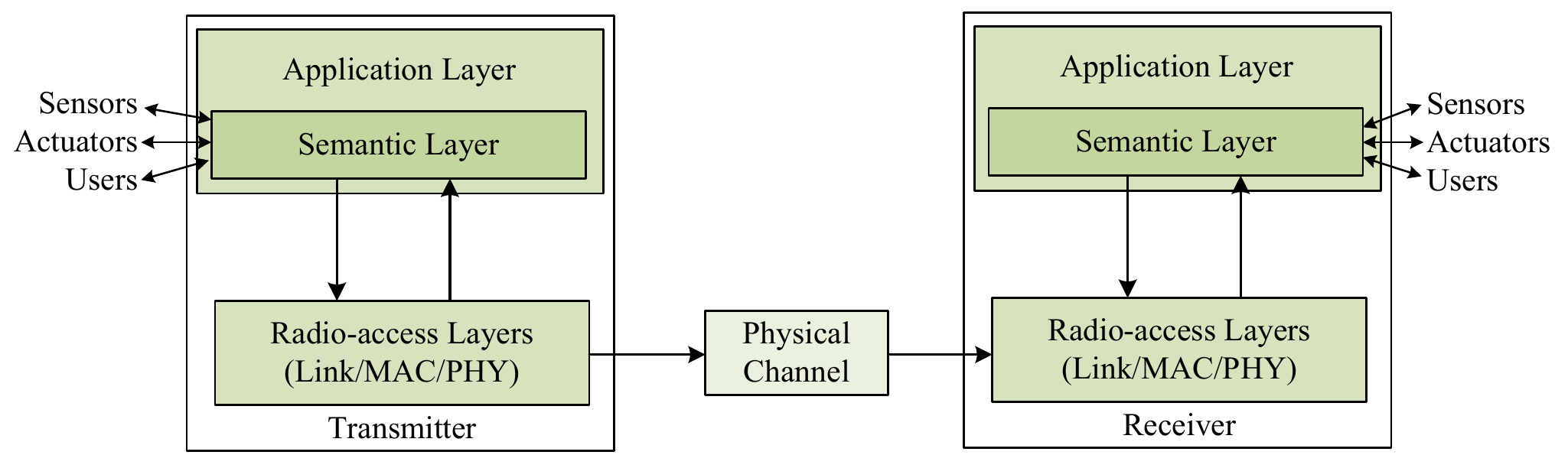}
\caption{The novel layered architecture of semantic communication system~\cite{Lan2021WhatIS}.}
\label{SemanticN}
\end{figure}

To support  semantic communications, a semantic Open System Interconnection (OSI) model was introduced~\cite{Lan2021WhatIS}. As shown in Fig.~\ref{SemanticN}, it allows the semantic layer to interface with sensors or actuators and to access algorithms and data for  specific tasks. In the semantic layer,  semantic coding is performed to transmits semantic encoded data to lower layers. Moreover, the radio-access layer aims at improving the system transmission performance, which transmits control signals to the semantic layer over a control channel. Those control signals are exploited by the semantic layer to remove semantic noise  for semantic symbol error correction or to control  computing at the application layer. Similarly, Zhang~\emph{et al.}~\cite{ZHANG2021} proposed a new semantic system model with different layers as a comprehensive system to replace the existing OSI model.

\subsection{Semantic Noise}
\begin{figure}[!t]
\centering
\includegraphics[width=4.0in]{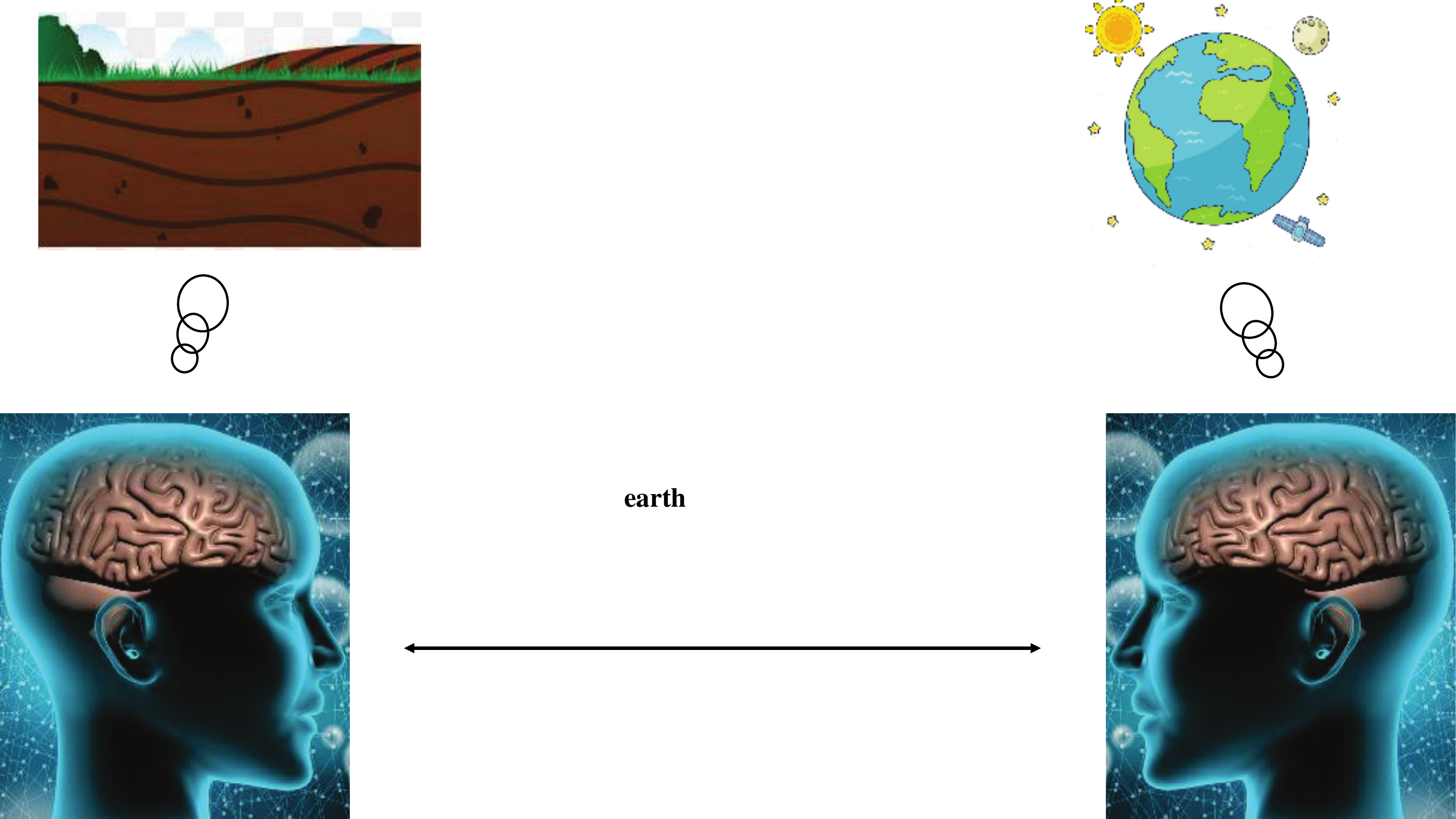}
\caption{An example of semantic mismatch for the word ``earth'' between two persons.}
\label{fig_semnoise}
\end{figure}

Semantic noise refers to the disturbance that affects the interpretation of the message, which could be treated as the semantic information mismatch between the transmitter and the receiver in semantic communications. Fig.~\ref{fig_semnoise} illustrates the semantic noise as the two parties interpret the word ``earth'' in different ways.

\begin{figure}[t]
\centering
\includegraphics[width=6.5in]{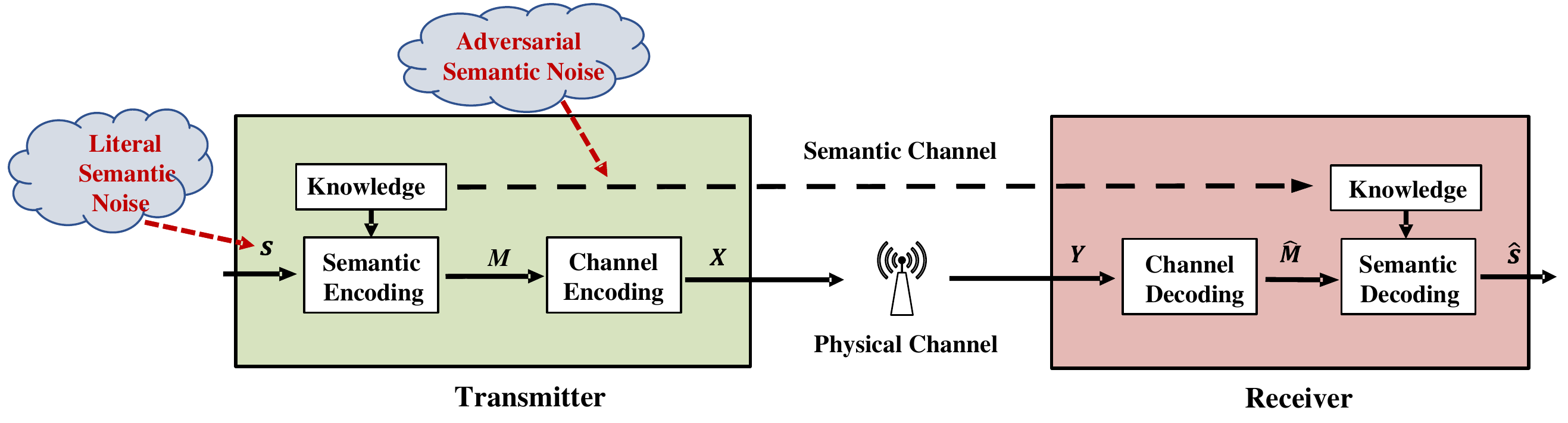}
\caption{Semantic noise in a semantic communication system.}
\label{semantic_noisy}
\end{figure}

There is yet a general formulation of semantic noise. For semantic communications, we categorize it into two types as shown in Fig.~\ref{semantic_noisy}. The first type of semantic noise refers to semantic ambiguity introduced to the source. For example, minor changes to letters or words in the sentence, for instance, replace the synonym  or reverse the alphabetical order randomly, could make it hard to understand the semantic meaning by the machine, which leads to wrong decision making~\cite{adv2}. In particular, Peng~\textit{et al.}~\cite{peng2022robust} developed a robust semantic communication system for text transmission to deal with such type of semantic noise. 

\begin{figure}[t]
\centering
\includegraphics[width=30pc]{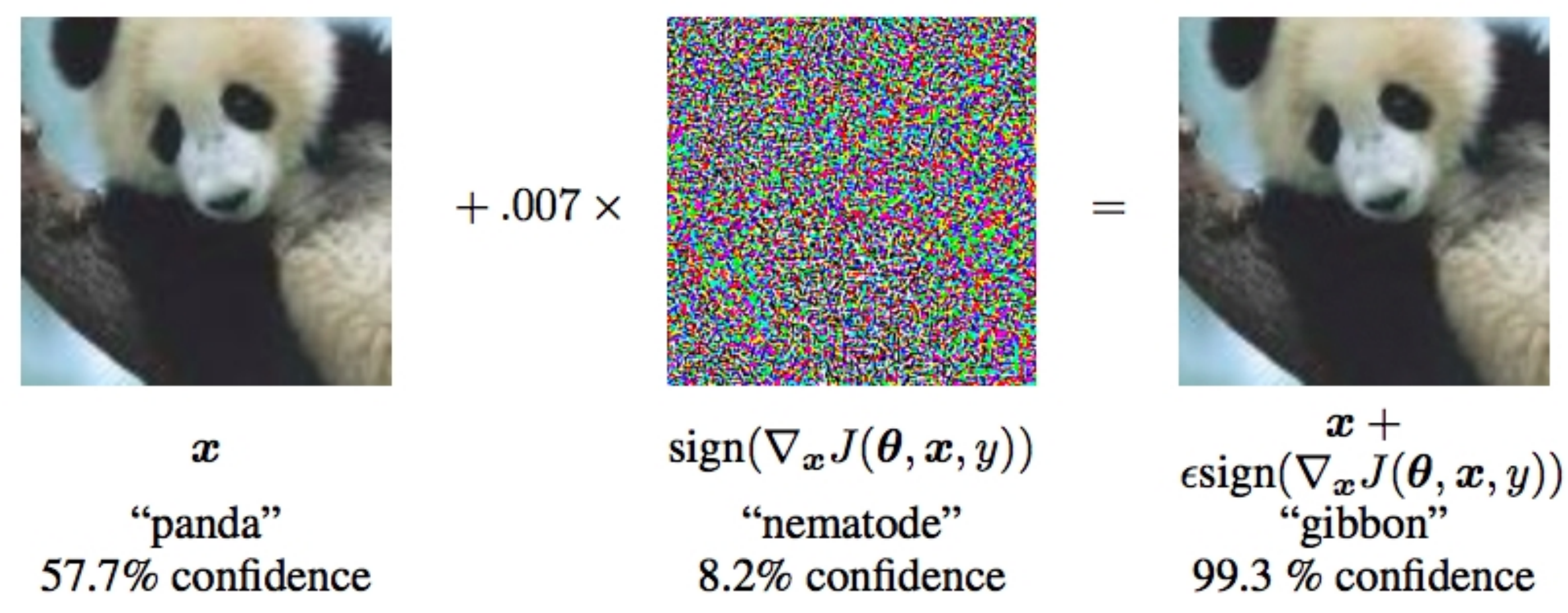}
\caption{An example of the adversarial example in image domain~\cite{image_adv_1}.}
\label{fig_adv_2}
\end{figure}

Another type is the adversarial semantic noise that misleads the DL model. Due to the discrete nature of  text, it is impossible to add perturbation to  text without being noticed by  human. However, some modifications added to images are so subtle that human can hardly notice. A typical example of adversarial example in the image domain is shown in Fig.~\ref{fig_adv_2}, in which the adversarial samples are added. We could see that the image with adversarial noise will mislead the DL models for classification but look the same as the original image if observed by human. Moreover, the sample-dependent and sample-independent semantic noise are investigated in~\cite{hu2022robust}, which both mislead the DL model.

Some prior works have studied adversarial example generation. Goodfellow~\textit{et al.}~\cite{image_adv_1}  proposed a fast gradient sign method to generate perturbation by using the gradient of the loss function. Miyato~~\textit{et al.}~\cite{adv1} developed a fast gradient method to generate adversarial examples. The role of adversarial examples for DL has two effects. First, it can be used to prevent machine learning system from attacks. Secondly, it is beneficial to improve the robustness of a DL-based system. Note that all the aforementioned adversarial examples are generated by human. To investigate whether there are adversarial samples in nature,  a mobile phone camera is used to photograph adversarial samples~\cite{Adversarial2017}, which have showed that images obtained by taking pictures of the adversarial samples will also be misclassified.

\subsection{Performance Metrics}
In  the conventional communication systems,  BER and SER are usually adopted as the performance metrics. However, they are not applicable to measure semantic communication systems any more as the focus of communications is shifted from accurate symbol transmission to effective semantic information exchange. At the  moment, a  general metric is still missing for semantic communications. In the following of this section, We will discuss  metrics for different sources, including text, image, and speech,  in the literature.

\subsubsection{\bf{Text Semantic Similarity}}
Word-error rate (WER) is used to measure the semantic text transmission~\cite{jiang2021deep,gold2018}, which is not applicable for semantic text transmission as  two sentences with different words may share high semantic similarity. The bilingual evaluation understudy (BLEU) score is a commonly used metric to measure the quality of text after machine translation~\cite{papineni2002bleu}, which has been exploited to measure semantic communication systems for text transmission~\cite{xie2020deep,xie2020lite}.  The BLEU score between the transmitted sentence $\bf s$  and the received sentence $\bf \hat s$ is calculated by
\begin{equation}
  \log {\text{BLEU = min}}\left( {1 - \frac{l_{\bf \hat s}}{l_{\bf s}},0} \right) + \sum\limits_{n = 1}^N {{u_n}\log {P_n}}, 
\end{equation}
where  $l_{\bf s}$ and $l_ {\bf \hat s}$ are the word length of $s$ and $\hat s$, respectively, $u_n$ defines the weights of the $n$-grams, and $P_n$ is the $n$-grams score defined as
\begin{equation}\label{wer}
{P_n} = \frac{{\sum\nolimits_k {\min \left( {{C_k}\left( {{\mathbf{\hat s}}} \right),{C_k}\left( {\mathbf{s}} \right)} \right)} }}{{\sum\nolimits_k {\min \left( {{C_k}\left( {{\mathbf{\hat s}}} \right)} \right)} }},
\end{equation}
where $C_k(\cdot)$ is the frequency count function for the $k$-th element in the $n$-th gram. BELU score counts the difference of $n$-grams between two sentences, where $n$-grams refer to the number of words in a word group for comparison. For example, for sentence ``This is a dog", the word groups are ``this", ``is", ``a" and ``dog" for $1$-gram. For $2$-grams, the word groups include ``this is", ``is a" and ``a dog". The same rule applies for the rest. 

The range of BLEU score falls between 0 and 1. The higher the score, the higher similarity between the two sentences. However, few human translations will attain the score of 1 since sentences with different expressions or words may refer to the same meaning. For example, sentences ``my bicycle was stolen" and ``my bike was stolen" share same meaning but the BLEU score is not 1 since they are different when compared word by word. 

To characterize such a feature, sentence similarity~\cite{xie2020deep} was proposed as a new metric to measure the semantic similarity level of two sentences, which is expressed as
\begin{equation}\label{ss}
   {\tau}\left( {\bf \hat s,s} \right) =  \frac{{{\bm B_{\bm \Phi} }\left( {\bf{s}} \right) \cdot {\bm B_{\bm \Phi} }{{\left( {{\bf{\hat s}}} \right)}^T}}}{{\left\| {{\bm B_{\bm \Phi} }\left( {\bf{s}} \right)} \right\|\left\| {\bm {B_{\bm \Phi} }\left( {{\bf{\hat s}}} \right)} \right\|}},
\end{equation}
where $ {\bm B_{\bm \Phi} }$ is the BERT model~\cite{DevlinCLT19} to map a sentence to its semantic vector space, which is a  pre-trained model with billions of sentences.  Instead of comparing  two sentences directly, we compare their semantic vectors obtained by the BERT model. Sentence similarity ranges from 0 to 1. The higher the value, the higher similarity between the two sentences.  

To achieve a tradeoff between the transmission accuracy and the number of symbols used for each message, a metric~\cite{sana2021learning} has been designed as
\begin{equation}
    \gamma  = \frac{1}{N} \times \left( {1 - \psi \left( {{\bf s},\hat {\bf s}} \right)} \right),
    \label{METRIC3}
\end{equation}
where $N$ is the number of symbols per message and ${\psi \left( {{\bf s},\hat {\bf s}} \right)}$ is the semantic error between ${\bf s}$ and  $\hat {\bf s}$. Note that ${\psi \left( {{\bf s},\hat {\bf s}} \right)}$ could be in various formats, such as BLEU score and mean-squared error (MSE), which is dependent on the transmission task at the receiver. 

Moreover, some other metrics have also been introduced  recently, i.e., average bit consumption per sentence  measures the system from a communication perspective~\cite{jiang2021deep}. 

\subsubsection{\bf{Image Semantic Similarity}}
The similarity of two images, $A$ and $B$, is  measured as
\begin{equation}
    \eta(f(A),f(B))=\|f(A)-f(B)\|_2^2,
\end{equation}
where $f(\cdot)$ is the image embedding function mapping an image to a point in the Euclidean space. The image embedding function $f(\cdot)$ is the essential part for finding the image semantic similarity. Note that the commonly used metrics, such as peak signal-to-noise ratio (PSNR)  and structural similarity index (SSIM), are  shallow functions and fail to count many nuances of human perception. Moreover, traditional image similarity metrics are built on the top of hand designed features, such as Gabor filter and scale-invariant feature transform (SIFT). Their performance is heavily limited by the representation power of  features.

Image semantic similarity metric depends on the high-order image structure, which is usually context dependent. DL-based image similarity metrics could achieve promising results, as the convolutional neural network (CNN) encodes high invariance and captures  image semantics. It is discovered that deep CNNs trained on a high-level image classification task are often remarkably useful as a representational space. For example, we can measure the distance of two images in VGG feature space as  the perceptual loss for image regression problems~\cite{johnson2016perceptual}. They define two perceptual loss functions based on a  VGG network, $\bm{\phi}$. The feature reconstruction loss encourages the two images with similar feature representations computed by  $\bm{\phi}$. Let $\bm{\phi}_i(x)$ be the  activation function of the $l$th layer, which is of shape $L$. The feature reconstruction loss is calculated by
\begin{equation}
    \mathcal{L}_\text{feature}^{\bm{\phi},l}(A,B)=\frac{1}{L}\|\bm{\phi}(A)-\bm{\phi}(B)\|_2^2.
\end{equation}
The style reconstruction loss penalizes differences in colors, textures, and common patterns. It is the  difference between the Gram matrices, $G_l^{\bm{\phi}}$, of the two images given by
\begin{equation}
    \mathcal{L}_\text{style}^{\bm{\phi},l}(A,B)=\|G_l^{\bm{\phi}}(A)-G_l^{\bm{\phi}}(B)\|_F^2.
\end{equation}

The effectiveness of deep features in similarity measuring is not restrict to VGG architecture. Richard~\emph{et~al.}~\cite{zhang2018unreasonable}  evaluated deep features across different architectures and tasks, which showed significant performance gain compared to all previous metrics and coincided with human perception. Furthermore, a deep ranking model  proposed in~\cite{wang2014learning}  characterizes the image similarity relationship within a set of triplets: a query image, a positive image, and a negative image. The image similarity relationship is characterized by the relative similarity ordering in   triplets. Moreover, metrics, including adversarial loss~\cite{goodfellow2014generative}, inception score (IS), and Fréchet  inception distance (FID) \cite{salimans2016improved}, have  also been proposed to measure the similarity between  images generated from generative adversarial networks (GAN) and natural images, from a image distribution perspective.

Visual semantic embedding is another way to assess the image semantic similarity \cite{frome2013devise}. As mentioned earlier, the  concepts from different images can be extracted and compared. The visual translation embedding (VTransE) network \cite{zhang2017visual} maps the visual features of objects and predication into a low-dimensional semantic space. Therefore, the semantic similarity can be measured. The relationship detection model in~\cite{zhang2019large} learns a module to map features from the vision and semantic modalities into a shared space, where the matched feature pairs should be discriminative against those unmatched ones and maintaining close distances to semantically similar ones. As a result, the model can  represent semantic similarity of images well at the relationship level. Semantic embedding has been widely used in scene graph generation (SGG), image captioning, and image retrieval. Hence, it is with great potential to be exploited in semantic communication systems.

\subsubsection{\bf{Speech Quality Measurement}}
The transmission goal of semantic commutations can be categorized into full data reconstruction and task execution. To achieve the speech reconstruction, the global speech semantic information, such as the voice of a speaker, text information, and speech delay, are transmitted and recovered at the receiver. Therefore, the metrics, such as perceptual evaluation of speech quality (PESQ)~\cite{rix2001perceptual}, short-time objective intelligibility (STOI)~\cite{5713237}, and perceptual objective listening quality assessment (POLQA)~\cite{beerends2013perceptual}, can be adopted to measure the global semantic content of speech signals, which comprehensively evaluate the reconstructed speech signals. 
In~\cite{Wen2021JSAC}, PESQ is adopted as the performance metric in a semantic communication system for speech transmission.

However, to serve intelligent tasks, i.e., speech synthesis, the speech signals are synthesized at the receiver based on the text  and speaker's information, which omits some content of semantic information, e.g., speech delay. Therefore,  unconditional Fréchet deep speech distance (FDSD) and unconditional kernel deep speech distance (KDSD) are utilized to assess the quality of synthesized speech, which first extracts the features of the speech signals and feeds these features into an assessment model to measure their similarity.

Denote the extracted features of the original speech samples and the synthesized ones as $\boldsymbol D\in\mathfrak R^{K\times L}$ and $\widehat{\boldsymbol D}\in\mathfrak R^{\widehat K\times L}$, respectively, the FDSD is defined as
\begin{equation}
\begin{split}
    \mathrm{\Gamma}^2=\left\|{\boldsymbol\mu}_{\boldsymbol D}-{\boldsymbol\mu}_{\widehat{\boldsymbol D}}\right\|^2   +\mathrm{Tr}\left[{\boldsymbol\Sigma}_{\boldsymbol D}+{\boldsymbol\Sigma}_{\widehat{\boldsymbol D}}-\left({\boldsymbol\Sigma}_{\boldsymbol D}{\boldsymbol\Sigma}_{\widehat{\boldsymbol D}}\right)^\frac12\right],
\end{split}
\label{cFDSD}
\end{equation}
where ${\boldsymbol\mu}_{\boldsymbol D}$ and ${\boldsymbol\mu}_{\widehat{\boldsymbol D}}$ represent the averages of $\boldsymbol D$ and $\widehat{\boldsymbol D}$, respectively, while ${\boldsymbol\Sigma}_{\boldsymbol D}$ and ${\boldsymbol\Sigma}_{\widehat{\boldsymbol D}}$ denote their covariance matrices.

KDSD~\cite{binkowski2019high} is given by
\begin{equation}
\begin{split}
    \mathrm{\Delta}^2=&\frac1{K\left(K-1\right)}\sum_{\underset{i\neq j}{1 \leq i,j\leq K}}q\left({\boldsymbol D}_{i,}{\widehat{\boldsymbol D}}_j\right)\\ 
    &+\frac1{\widehat K\left(\widehat K-1\right)}\sum_{\underset{i\neq j}{1 \leq i,j\leq\widehat K}}q\left({\boldsymbol D}_{i,}{\widehat{\boldsymbol D}}_j\right) 
    +\sum_{i=1}^K\sum_{j=1}^{\widehat K}q\left({\boldsymbol D}_{i,}{\widehat{\boldsymbol D}}_j\right),
\end{split}
\label{cKDSD}
\end{equation}
where $q(\cdot)$ is the kernel function.

From the above,  for semantic communications serving different tasks, the performance metric is heavily dependent on the chosen ``semantic language'' for the application. Such a semantic language could be a typical natural language, the scene graph for image processing, truth tables from Carnap~\emph{et al.}~\cite{carnap1952outline}, or the customized graph-based language from Kalfa~\emph{et al.}~\cite{KALFA2021103134}. The metrics will have to be adapted to these languages.

\section{Deep  Semantic Communications for Text, Speech, and Image/Video}
Though  semantic theory has been investigated for decades, the lack of a general mathematical tool limits its applications. Thanks to the advancements of DL, some interesting works have been developed for semantic communications  in recent years. This section  presents the latest work on the deep semantic communication system design for text, image, speech, and multimodal data.

\subsection{Text Processing}
The advancement on NLP~\cite{otter2020survey} enables  text coding to consider the semantic meaning of text, which motivates us to re-design the transceiver for achieving successful  semantic information transmission. For the system  shown in Fig.~\ref{fig_semsys}, neural networks are  used to represent the  transmitter and receiver in DL-enabled semantic communications. The core of DL-enabled semantic communication is to design semantic coding, which can understand and extract semantic information. Those semantic features are then compressed. Moreover, the channel decoder is trained to combat the channel impairment. Another core task is to design a proper loss function to minimize semantic errors and channel impairments. If the system is designed for serving a specific task at the receiver, the loss function should be adjusted accordingly to capture features relevant to the task.

For erasure channels,  Farsad~\textit{et al.}~\cite{gold2018} developed a long short-term memory (LSTM) enabled JSCC for text transmission. The cross entropy is adopted as the loss function and WER is used for performance metric. It shows the great potential of DL-enabled JSCC compared to the conventional communication system. Though the concept of semantic communication was not mentioned in~\cite{gold2018}, it inspired subsequent research greatly. 

\begin{figure*}[tbp]
\begin{minipage}[t]{0.5\linewidth}
\centering
\includegraphics[width=1.2\textwidth]{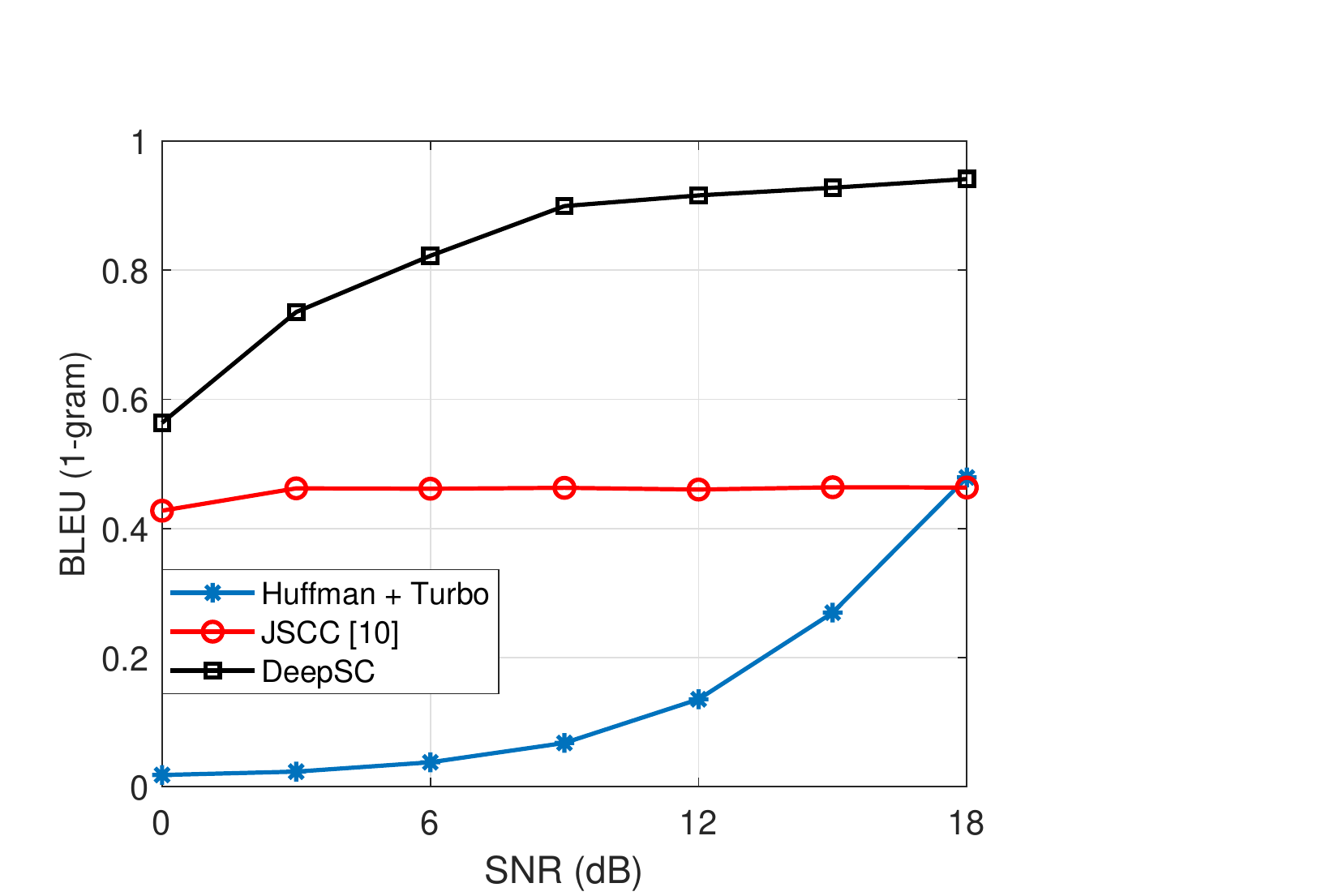} 
\end{minipage}
\begin{minipage}[t]{0.51\linewidth}
\centering
\includegraphics[width=1.2\textwidth]{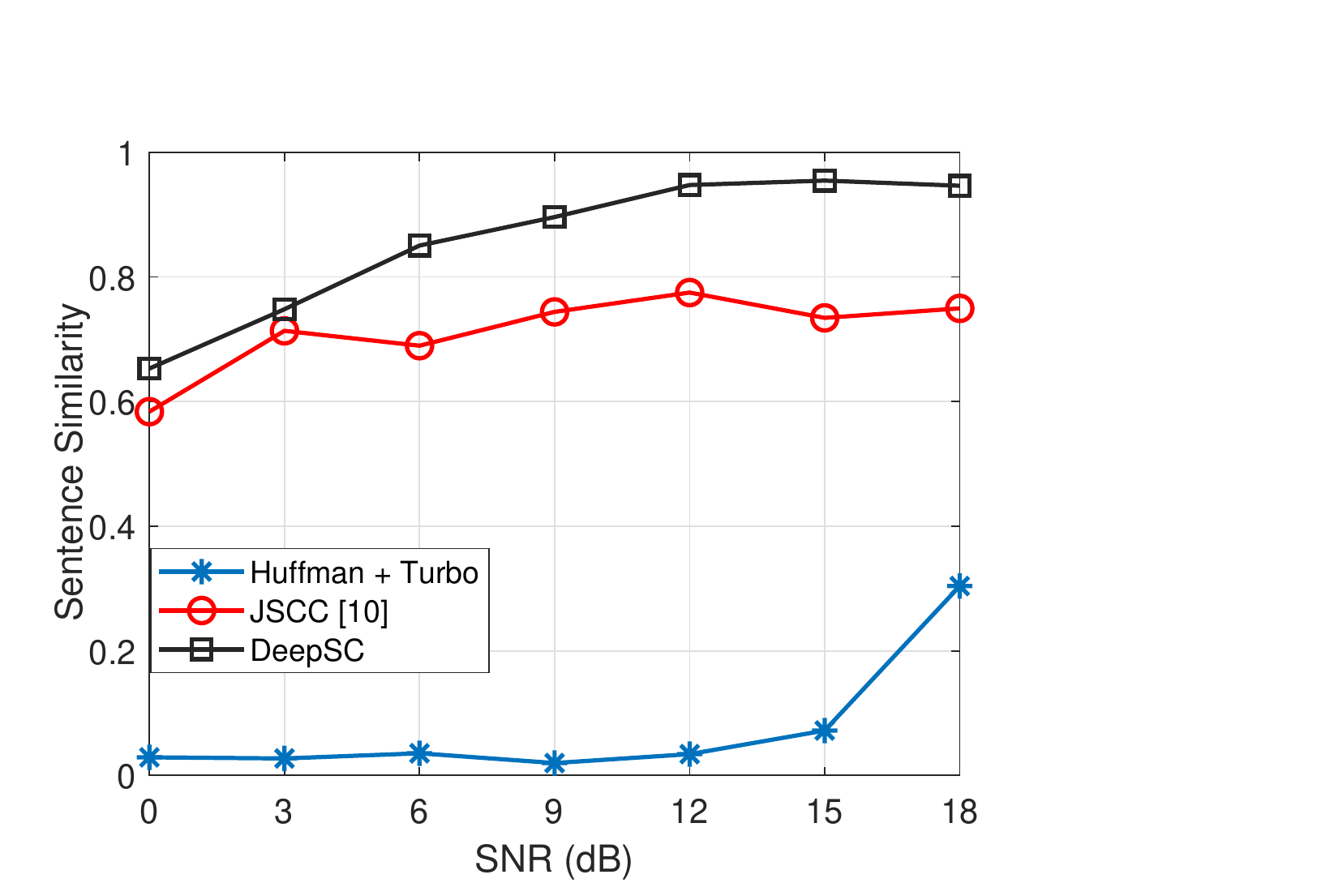}
\end{minipage} 
\caption{BLEU score comparison of DeepSC and traditional methods under Rayleigh fading channels, with 8 symbols  used for each word \cite{xie2020deep}.}
\label{DEEPSC}
\end{figure*}

Afterwards, Xie~\textit{et al.}~\cite{xie2020deep} developed a Transformer based joint semantic-channel coding for an end-to-end semantic communication system, named DeepSC. In such an end-to-end system, the block structure of the conventional communication system has been merged~\cite{QINDL:2019}. In particular, a new loss function given by~\eqref{Xie_loss} was provided, which counts the cross entropy for better understanding texts and the mutual information for higher data rates~\cite{xie2020deep}. By designing such semantic coding and channel coding layers, DeepSC is capable of extracting  semantic features and guaranteeing their accurate transmission. It has been verified that DeepSC outperforms typical communication systems significantly, especially when channel conditions are poor. For instance, the BLEU score in Fig. \ref{DEEPSC} is improved by 800\% compared to the conventional method with Huffman coding and Turbo codes when SNR = 9~dB. When measured by sentence similarity defined in~\eqref{Xie_loss}, we could also see the significant performance gap between DeepSC and the conventional method. Note that when SNR = 12 dB, the BLEU score is less than 0.2, which makes the received sentence almost impossible to read by human beings. It could be reflected by the  sentence similarity which is almost 0 with SNR = 12 dB.

Moreover, Table \ref{table 333} provides an snapshot of the received sentences after they pass through Rayleigh fading channels, in which the traditional methods exhibit some spelling mistakes. Note that the performance metrics, such as BLEU score and sentence similarity, cannot be used as the loss function in DeepSC as it will cause gradient disappearance for the transceiver training.

\begin{table*}[t]
\small
\caption{An comparison of received sentences with DeepSC and benchmarks when transmitted over Rayleigh fading channels, SNR=18 dB, and 8 symbols  used for each word.~\cite{xie2020deep}.}
\label{table 333}
\centering
\begin{tabular}{ |c| c |} 
\hline
Transmitted sentence & it is an important step towards equal rights for all passengers. \\
\hline
DeepSC &  it is an important step towards equal rights for all passengers.  \\
\hline
JSCC\cite{gold2018} & it is an essential way towards our principles for democracy.  \\
\hline
Huffman + Turbo  &  rt is a imeomant step  tomdrt equal  rights for atp passurerrs.  \\
\hline
\end{tabular}
\end{table*}

Since then, several variants of DeepSC have been developed. Specifically, Jiang \textit{et al.}~\cite{jiang2021deep} combines semantic coding with Reed Solomon coding and hybrid automatic repeat request for improving the reliability of  text semantic transmission. A similarity detection network was proposed to detect meaning errors. Moreover, Sana~\textit{et al.}~\cite{sana2021learning} defined a new loss function  \eqref{Loss3} to capture the effect of semantic distortion, which can dynamically trade semantic compression loss for semantic fidelity. Note that \eqref{Loss3} is upper bounded by \eqref{Xie_loss}, but it is hard to use \eqref{Loss3} as the loss function for  neural network training. Another contribution from ~\cite{sana2021learning} is the introduction of adaptive number of symbols used for each word   for further performance gain in terms of the metric defined in~\eqref{METRIC3}. The performance gain becomes obvious when the maximum number of symbols for each word is large, but it also increases the size of data to be transmitted. To make the trained model affordable for capacity-limited IoT devices, Xie~\textit{et al.}~\cite{xie2020lite} developed a lite model of DeepSC by  pruning and quantizing  the trained DeepSC models, which could achieve 40x compression ratio without performance degradation.

\subsection{Image Processing}

In this part, we first introduce  different approaches of image semantic extraction. DL-based image compression and semantic communications for image transmission are then discussed.

\subsubsection{\bf{Non-Structural Image Semantic Representation}}
The representation of pixel-level images usually lacks of high-level semantic information. Classical machine learning methods utilize hand-crafted features for image representation. Later on, sparse coding~\cite{bristow2013fast} was introduced  to represent image patches as a combination of overcomplete basis elements, also known as a codebook. However, the representation power of  shallow features is usually limited.

After the breakthrough of convolutional neural networks (CNN), powerful deep features become available. In a CNN, each layer generates a successively higher-level abstraction of the input data, named a feature map, to keep essential yet unique information. By employing a very deep hierarchy of layers, modern CNNs achieve superior performance in image semantic representation and content understanding. Although these approaches have achieved advanced performance in visual feature extraction, there is still a gap between visual features and  semantics.

To narrow the gap, some researchers focus on extracting   image utilizing context information. The deep semantic feature matching approach proposed in~\cite{ufer2017deep}  incorporates convolutional feature pyramids and activation guided feature selection. Promising results have been obtained in estimating correspondences across different instances and scenes of the same semantic category. Huang~\cite{huang2018learning} proposed an image and sentence matching approach by utilizing the image global context  to learn  semantic concepts, such as objects, properties, and actions. The description of image regions is generated in \cite{karpathy2015deep}, where the visual-semantic alignment model infers the alignment between segments of sentences and the region of the image described by those text segments. Shi~\textit{et~al.}~\cite{shi2015transferring} proposed a semantic representation for person re-identification, where  semantic attributes include the color and category of clothes as well as different parts of the body. 

Some researchers tried to extract concise representation for image semantics, such as semantic segmentation map, sketch, object skeleton, and facial landmark. These semantic labels describe the layout of those objects. For instance, each pixel in semantic segmentation is labeled with the class of its enclosing object or region. Object skeleton describes the symmetry axis, which is widely used in object recognition/detection. Facial landmarks are used to localize and represent salient regions of the face, i.e., eyes and nose, which are applied to various tasks, such as  face alignment, head pose estimation, and face swapping, to name a few.

\subsubsection{\bf{Structural Image Semantic Representation}}
The efforts of structural image semantic representation have been concentrated on using graph-based machine learning techniques to reduce  semantic gap  between the low-level image features and the profusion of high-level human perception to images. With the power of graph representation, the solution space is reduced, which results in faster optimization convergence and higher accuracy in the representation learning. Graph based methods have demonstrated effective performance in image segmentation, annotation, and retrieval. In this way, the gap between image vision content and the semantic tags can be bridged.

\begin{figure}[t]
    \centering
    \includegraphics[width=0.65\linewidth]{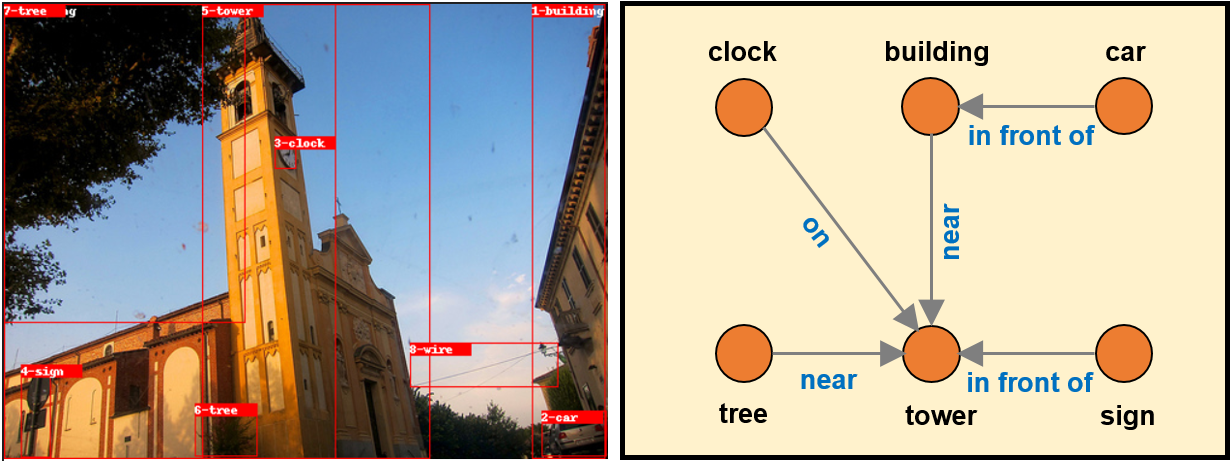}\
    \caption{A scene graph example.}
    \label{SGGexample}
\end{figure}

Scene graph \cite{johnson2015image} has been regarded as a typical representation of the semantic graph, which is a data structure to describe objects and their relationship in a real scene. As shown in Fig.~\ref{SGGexample}, a complete scene graph could represent the detailed semantics of scenes, which can be used to encode 2D/3D images/videos into  semantic features. Scene graph is a new content for scene description, which  has been widely adopted in inference tasks, such as question answering, image retrieval, and image captioning.

Scene Graph Generation (SGG) is developed to build a more complete scene graph. Particularly, VTransE places objects in a low-dimension relation space where the relationship can be modeled as a simple translation vector ($subject+predicate=object$) \cite{zhang2017visual}. Note that elements of visual scenes have strong structural regularity, thus some structural repetitions in scene graphs can be examined. MOTIFNET \cite{zellers2018neural} divides SGG into the stage predicting bounding boxes, object labels, and relationships. Moreover, Causal-TDE \cite{tang2020unbiased}, which is based on causal inference other than the conventional likelihood, introduced a causal graph of SGG to remove the effect from the undesired bias by counterfactual causality.

\subsubsection{\bf{Deep Learning based Image Compression}}

 Traditional image compression  projects an image into its sparse domain and reconstructs the image under the guidance of pixel level accuracy. They suffer from blocky and ringing artifact at low bit rate and the reconstructed image is not visually pleasant. It is the frontier that machine learning based image compression methods transform the traditional  pixel reconstruction into semantic reconstruction. Content understanding is considered as the core of  next generation image coding. 
 
 Deep autoencoder encodes the image into a  low dimensional latent code thus achieves highly efficient compression. Various end-to-end deep autoencoder based image compression architectures were proposed in \cite{theis2017lossy, cheng2018deep, balle2016end}. To deal with the non-differential rounding based quantization,  differentiable alternatives have been proposed for the quantization and entropy rate estimation. Based on the fact that local information content of an image is spatially variant, a content-aware bit rate allocation method was proposed in \cite{li2018learning}. 
 
 Moreover, GAN is used to produce visually pleasing reconstruction for very low bit rates. The pioneering work~\cite{rippel2017real}  introduced GAN to compression and proposed a real-time adaptive image compression method. By utilizing an autoencoder feature pyramid to downsample an image and a generator to reconstruct it, the compressor typically produces files 2.5 times smaller than the typical image compression method, such as JPEG. Agustsson \textit{et al.}~\cite{agustsson2019generative} introduced the GAN based extreme learned image compression, which obtains visually pleasing results at an extremely low bit rate. Furthermore, if a semantic label map is available, the non-essential regions in the decoded image can be fully synthesized. Wu~\textit{et al.}~\cite{wu2020gan} proposed a GAN-based tunable image compression system, in which the important map is learned to guide the bit allocation.

\subsubsection{\bf{Semantic Communications for Image/Video}}
\begin{figure}
    \centering
    \includegraphics[width=5.5in]{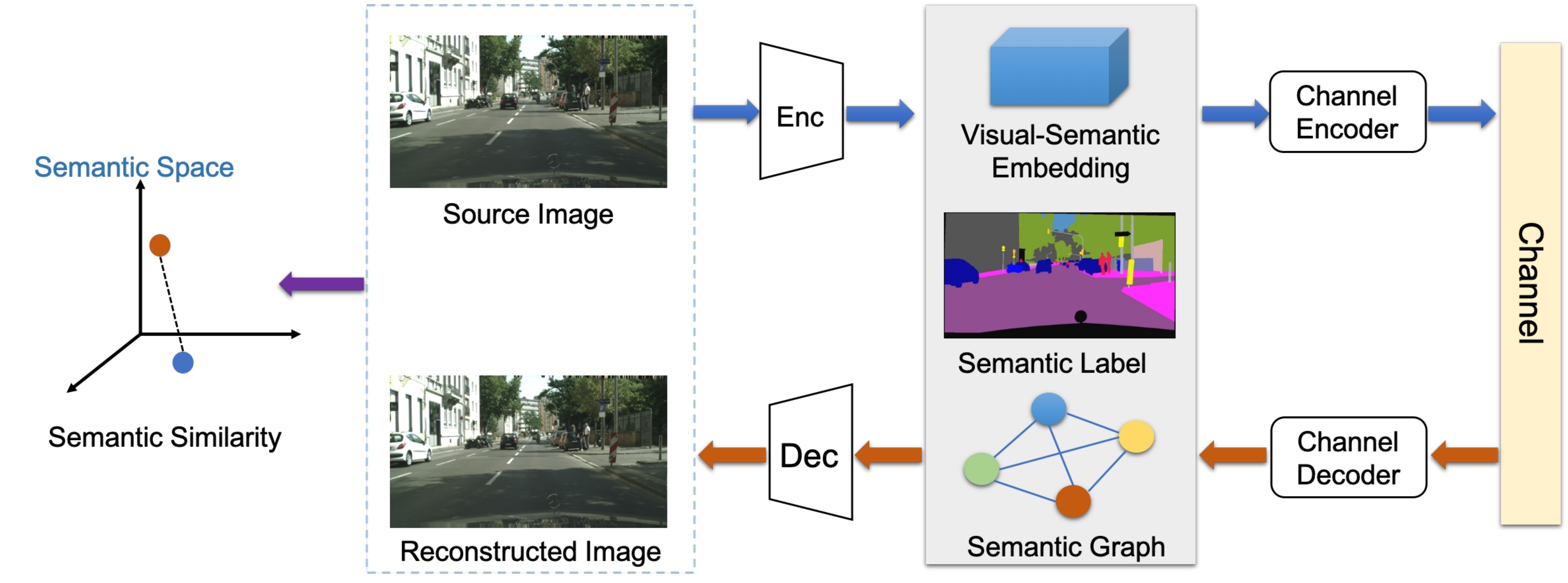}
    \caption{A semantic communication framework for image transmission.}
    \label{fig:framework}
\end{figure}

Fig. \ref{fig:framework} illustrates a structure for image semantic communications, which utilizes the aforementioned image processing but is beyond that. The semantic encoder extracts the low-dimensional semantic information, in the form of visual-semantic embedding, semantic label, or semantic graph. Machine learning techniques are adopted to design an efficient semantic encoder. The next stage is the channel encoder, which can be jointly trained with the semantic encoder.  

Particularly, Kurka~\textit{et al.} proposed DeepJSCC \cite{kurka2020deepjscc,kurka2020bandwidth} for adaptive-bandwidth wireless transmission of images. It  exploits the channel output feedback signal and outperfoms separation-based schemes. DeepJSCC performs well in the  low SNR and small bandwidth regimes with slight degradation. At the receiver, the semantic decoder usually adopts a GAN based architecture to map  semantic information into its vision space. Such an under-determined image reconstruction training task is optimized using the criterion of semantic similarity. 

More works have been designed for serving certain vision tasks, named task-oriented semantic communications. Specifically, Lee~\textit{et al.}~\cite{Lee:2019} designed a joint transmission-classification system for images, in which the receiver outputs image classification results directly. It has been verified that such a joint design achieved higher  classification accuracy than performing image recovery and classification separately. Kang~\textit{et al.}~\cite{kang2021taskoriented}  proposed a scheme for joint image transmission and scene classification. Deep reinforcement learning was exploited to identify the most essential semantic features for serving the transmission task so as to achieve the best tradeoff between   classification accuracy and transmission cost. Jankowski~\textit{et al.}~\cite{jsac-JankowskiGM21} considered image based re-identification for persons or cars as the transmission task, and two schemes were proposed to improve the retrieval accuracy.   

Moreover, video transmission is considered as a killing application of semantic communications, especially the video conference. In particular,  Wang~\textit{et al.}~\cite{tung2021deepwive} have developed a reinforcement learning enabled end-to-end framework for video transmission with variable bandwidth. It demonstrates superior performance compared to the conventional method.  Wang~\textit{et al.}~\cite{wang2022wireless} have designed a JSCC scheme for video transmission over the air with to minimize the end-to-end transmission rate-distortion. Jiang~\textit{et al.}~\cite{jiang2022wireless} have proposed a semantic transmission scheme for video conferencing with a novel semantic error detector. The photo of the speaker is shared as prior information to help reconstruct the motion of the speaker's facial expression. The developed scheme lower the demands on wireless resource dramatically. Tao~\textit{et al.}~\cite{Tao:2019} developed a mobile video transmission framework to guarantee the quality of experience (QoE). A large dataset has been built to find the relationship between the subjective QoE scores and the neural network parameters to guide the semantic video transmission. Moreover, Fried~\textit{et al.}~\cite{fried2019text} have proposed to edit talking-head video by editing text. Afterwards, Tandon~\textit{et al.}~\cite{tandon2021txt2vid} have proposed to transmit text only rather than video, which dramatically lower the network traffic.

For  image or video transmission, task-oriented semantic communications lower the network traffic significantly. However, as the system is trained for a specific task, the trained model should be updated or even re-trained if the transmission task varies.

\subsection{Speech and Multimodal Data Processing}
\subsubsection{\bf{Semantic Communications for Speech}} Semantic communication systems have also been designed for speech transmission~\cite{Weng2022,Tong:2021}. Weng~\textit{et al.}~\cite{Wen2021JSAC} proposed an extension of DeepSC for speech signals, named DeepSC-S. In particular,  joint semantic-channel coding  could deal with source distortion and channel effect. In this work, the bit-to-symbol transformation is not involved. MSE is adopted as the loss function to minimize the difference between the recovered speech signals and the input ones. In addition, signal-to-distortion ratio (SDR) and PESQ are adopted as two performance metrics to the quality of recovered speech signals. Tong~\textit{et al.}~\cite{Tong:2021} extended it to a multi-user case and implemented federated learning  to collaboratively train the CNN based encoder and decoder over multiple local devices and the server.  MSE is also used as the loss function and  performance metric, which hardly reflects the amount of  semantic information  at the receiver. 

Inspired by the unprecedented demands of intelligent tasks, DeepSC-ST~\cite{Weng2022} utilized  recurrent neural networks (RNNs) to extract the text-related semantic information from  speech signals at the transmitter and to recover the text sequence at the receiver for speech synthesis. By doing so, only text semantic information is transmitted, which reduce the required transmission resource significantly. Connectionist temporal classification (CTC)~\cite{graves2006connectionist} is adopted as the loss function, where character-error-rate (CER) and WER are adopted as two performance metrics to measure the accuracy of the recognized text information. The aforementioned FDSD and KDSD are exploited to measure the similarity between the real and  synthesized speech signals.

\subsubsection{\bf{Unified Semantic Communications for Multimodal Data and Multi-Task}}
\begin{figure}
    \centering
    \includegraphics[width=5.7in]{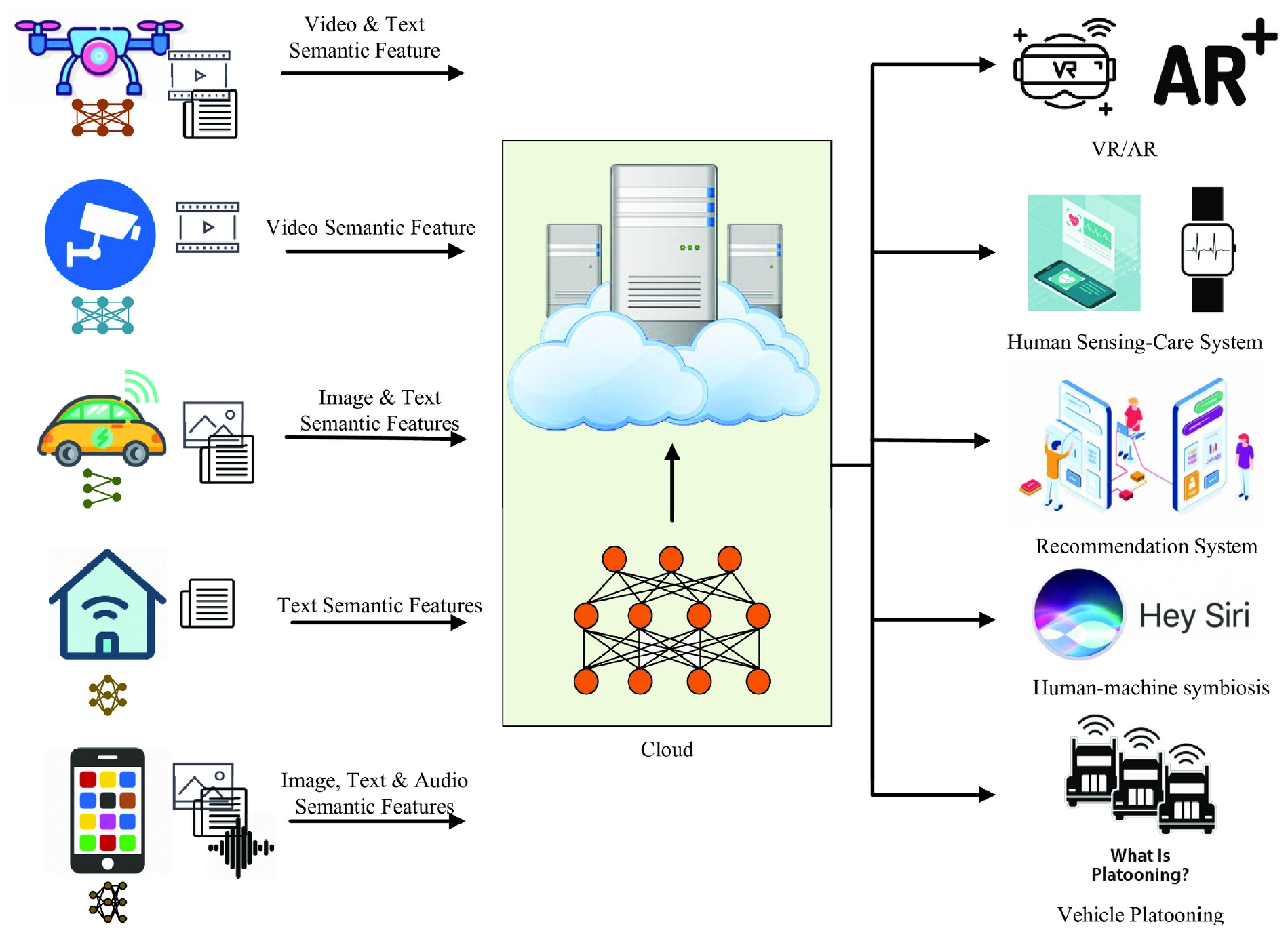}
    \caption{The illustration of semantic communication applications.}
    \label{multimodal}
\end{figure}

Multimodal data processing has been considered as a  critical task as shown in Fig.~\ref{multimodal}. For tasks,  such as AR/VR and human sensing care system,  the generated multimodal data  are correlated in the context. By introducing new degrees of freedom, multimodal data improves the performance of  intelligent tasks \cite{LahatAJ15}. 

Semantic communications are promising to support multimodal data transmission. Xie~\textit{et al.} \cite{xie2021taskoriented} developed  MU-DeepSC  for  the  visual question answering task, where text based questions about images are transmitted by one user and the enquiry images are transmitted from another user. Cross entropy is adopted as the loss function while the task related metric, i.e., answer accuracy rate, is used to measure the performance of MU-DeepSC. Different from all the aforementioned works for the point-to-point transmission, MU-DeepSC is designed for serving multi-user transmission. As an extension of DeepSC, a Transformer based  framework~\cite{Xie2022} has been developed as a unique structure for serving different tasks. Various tasks have been tested in~\cite{Xie2022} to show its superiority. 

Note that the aforementioned work still requires model training for each task, which limits the application. A unified deep learning enabled semantic communication system (U-DeepSC)~\cite{zhang2022unified} has been designed to serve various transmission tasks. To jointly serve these tasks in one model, domain adaptation is employed  to lower the transmission overhead. Moreover, since each task is with different difficulty and requires different numbers of layers, a multi-exit architecture has been proposed to provide early-exit results for relatively simple tasks.

Note that the  investigation on semantic commutations for multimodal data transmission is still at its infancy. But we could see the great potential of semantic communications to support multimodal data transmission for various applications, especially the great potential to lower the size of data to be transmitted by utilizing the correlations among data from different modalities.

\section{Research Challenges and Conclusions}
We can now conclude that the semantic communication is a breakthrough of the conventional communication. However, its general structure and many related issues are not clear yet, which motivates us to investigate more in this area.  To pave the  way to semantic communications, the following open questions should be answered:

\begin{enumerate}
    \item \textbf{Semantic theory}: Though some researchers have tackled  semantic theory in past decades, most of them are based on logical probability with limited application scenarios, which follows the framework of conventional information theory. It is still questionable whether we could follow a similar path to quantify semantic communications by semantic entropy, semantic channel capacity, semantic level rate-distortion theory, and the relationship between inference accuracy and transmission rate. 
    
    \item \textbf{Semantic transceiver}: Semantics provide concise and effective representations, thus making the semantic communication an efficient system in terms of bandwidth saving and subsequent task processing. However,  a general semantic level JSCC for different types of sources is not available yet. Moreover, it is a significant challenge to design a semantic noise robust communication system, incorporating  machine learning techniques like adversarial training. Furthermore,  proper loss functions without causing gradient disappearance are required. 
    \item \textbf{Semantic communications with reasoning}: Inspired by the advances of System 2 that enables reasoning, planning, and handling exceptions, a semantic communication system with reasoning could significantly reduces the communication cost by sending only the most effective semantics as pointed out by Tong's keynote and Seo~\textit{et al.} ~\cite{Seo:2021}. However, the investigated in this direction is still at its infancy, more efforts are expected to develop a more intelligent semantic communication system with reasoning.
    
    \item \textbf{Resource allocation in semantic-aware network}: In semantic-aware networks, it is essential to rethink  resource allocation for  semantic interference control. In contrast to  resource allocation in  conventional communications, which focuses on  engineering issues, i.e., improving bit transmission rate, semantic-aware resource allocation aims to address both  engineering  and semantic issues. The objective of semantic-aware resource allocation is to improve communication efficiency in semantic domain. In particular, semantic spectrum efficiency has been proposed in~\cite{yan2022resource,yan2022qoe}. However, this issue still faces following challenges: 
\begin{itemize}
    \item \textit{How to evaluate  semantic communication efficiency, i.e., semantic transmission rate or semantic spectral efficiency? }
    \item \textit{How to formulate a general resource allocation problem for difference task-oriented semantic systems to optimize the resource allocation policy to maximize  semantic communication efficiency? }
\end{itemize}

\item \textbf{Performance metrics}: Though several new performance metrics have been explored in semantic communication systems as aforementioned, it is more than desired to design more appropriate evaluation metrics for semantic communications, for instance, the metric to evaluate the amount of semantic information that has been preserved or missed. Moreover, a general performance metric, such as SER or BER for conventional communication systems, is required to measure different semantic communication systems.
\item \textbf{Applications}: Apart from the extensive research interest in semantic communications, the killing applications are more then desired. We are witnessing extensive interest in semantic communications enabled AR/VR and video conference from both academia and industries. We are looking forward to seeing more potential applications of semantic communications in the near future.

\end{enumerate}

\bibliographystyle{IEEEtran}
\bibliography{reference}

\end{document}